\documentclass[journal,12pt,onecolumn,letterpaper]{IEEEtran}
\usepackage{arxiv}
\usepackage{geometry}
\usepackage{times}
\usepackage{cite}
\usepackage{url}
\usepackage{graphicx}
\usepackage{lscape}
\usepackage{subfigure}
\usepackage{rotating}
\usepackage{rotfloat}
\usepackage{xcolor}
\usepackage{amsmath}
\usepackage{amssymb}
\usepackage[linesnumbered,ruled,vlined]{algorithm2e}
\usepackage{pseudocode}
\usepackage{tabularx}
\usepackage{caption}
\usepackage{array}
\usepackage[english]{babel}
\usepackage{gensymb}
\usepackage{textcomp}
\usepackage{placeins}
\usepackage{balance}
\usepackage{booktabs}

\title{TUSH-Key: Transferable User Secrets on Hardware Key\\
}

\author{Aditya Mitra \\
Centre of Excellence, Artificial Intelligence \& Robotics (AIR),\\
 School of Computer Science and Engineering\\
 VIT-AP University, India \\
\texttt{adityamitra5102@gmail.com}\\
\And
Anisha Ghosh \\
Centre of Excellence, Artificial Intelligence \& Robotics (AIR),\\
School of Computer Science and Engineering\\
VIT-AP University, India \\
\texttt{ghoshanisha2002@gmail.com}\\
\And
Sibi Chakkaravarthy Sethuraman \\
Centre of Excellence, Artificial Intelligence \& Robotics (AIR),\\
School of Computer Science and Engineering\\
VIT-AP University, India \\
\texttt{sb.sibi@gmail.com} \\
}

\begin{document}

\maketitle

\begin{abstract}
Passwordless authentication was first tested for seamless and secure merchant payments without the use of passwords or pins. It opened a whole new world of authentications giving up the former reliance on traditional passwords. It relied on the W3C Web Authentication (WebAuthn) and Client to Authenticator Protocol (CTAP) standards to use the public key cryptosystem to uniquely attest a user’s device and then their identity. These standards comprise of the FIDO authentication standard. As the popularity of passwordless is increasing, more and more users and service providers are adopting to it. However, the concept of device attestation makes it device-specific for a user. It makes it difficult for a user to switch devices. FIDO Passkeys were aimed at solving the same, synchronizing the private cryptographic keys across multiple devices so that the user can perform passwordless authentication even from devices not explicitly enrolled with the service provider. However, passkeys have certain drawbacks including that it uses proprietary end to end encryption algorithms, all keys pass through proprietary cloud provider, and it is usually not very seamless when dealing with cross-platform key synchronization. To deal with the problems and drawbacks of FIDO Passkeys, the paper proposes a novel private key management system for passwordless authentication called Transferable User Secret on Hardware Key (TUSH-Key). TUSH-Key allows cross-platform synchronization of devices for seamless passwordless logins with FIDO2 specifications.
\end{abstract}


\section{Introduction}

Passwords have been an issue since 1961 when it was invented by Fernando Corbató for the Compatible Time-Sharing System (CTSS) in MIT \cite{Fano_1966}. It was not long before 1966 when one of the graduate students, Allan Scherr caused the first password breach \cite{Fano_1966}. Since then, there has been several ways to protect passwords like hashing and encrypting but then again, there has been several ways to breach them, starting from brute forcing rainbow tables. Further, passwords falling in the memory-based authentication factor makes the humans a big attack vector \cite{Al-Kabir_2022}. They are prone to phishing and other social engineering-based attacks that can compromise passwords. To overcome the same, passwordless authentication has come into the picture with key fobs, smart cards, and other possession-based authentication methods.

Passwordless Authentications \cite{Huseynov_2022,KINGO_2023} have been on the rise with the wide adoption of FIDO2 specifications and has been acting as a replacement of traditional passwords \cite{Bicakci_2022}. Prior to FIDO2 specifications, there have been attempts at implementing passwordless authentication, but they were proprietary implementations and were not so widely adopted. FIDO2 implementation has been easier to adopt with the W3C WebAuthn and FIDO Alliance CTAP protocol being implemented in all common operating systems and web browsers. FIDO2 brings in the concept of device-attestation where the user’s identity is bound to his device. But this brings in the issue of multi-device authentication. It raises the concern that when a user owns multiple devices and needs to log in to a particular service provider using a different device that the one, he usually uses, how the service provider would identify the new device to be bound to his account.

The proposed standard proposes a seamless and secure key management system so that the user can authenticate from any of his device seamlessly. The contributions in this paper include:
\begin{itemize}
    \item A novel way of key management and synchronization.
    \item Keys will be stored in secure storage media like TEE/TPM only.
    \item Private keys will not be stored in proprietary cloud.
    \item Private keys will not be cloned on other devices.
\end{itemize}

\section{Background and Research Gap}

Entity authentication has been of utmost importance since the beginning of shared computing resources and time. It has taken various forms including passwords, personal identification numbers (PINs), biometrics, cryptographic authentication, device attestation and so on. User-based authentication were widely used in most information systems and most of them used Knowledge Based authentication (KBA), mainly passwords  \cite{Al-Kabir_2022, Laing_2022, Liou_2021}. However, it must be noted that passwords often have a wide range of vulnerabilities, most importantly the human factor. It is quite feasible to guess, reuse, phish and expose passwords. A few of the common ways have been finding smudge and thermal residue on shared devices \cite{Kaczmarek_Ozturk_Tsudik_2019}. Hence, the need for passwordless authentication \cite{Parmar_2022}. 

Device attestation \cite{Arias_2018} is a promising form of passwordless authentication and it ensures the operational demands for cyber-physical systems are met. Device attestation forms the foundation for cryptographic methods of passwordless authentication. Device attestation for user authentication binds the identity of a user to his devices. This simplifies user authentication and makes it more seamless. This lays the founding points of FIDO specifications \cite{Morii_2017}. FIDO Alliance gives a set of specifications that uses the W3C Web Authentication \cite{Hodges_2021} in conjunction with Client to Authenticator Protocol (CTAP) and is used to perform user authentication by device attestation. In 2022, with the announcement of passkeys \cite{Campbell_2023}, it has started to be widely used. Major Original Equipment Manufacturers (OEMs) have been providing native support to FIDO specifications. In earlier other research, FIDO specifications have been adopted for secure user-authentication on metaverse based environments \cite{Sibi_2022} and for hardware assets including in cyber-physical systems \cite{Sibi_Loki_2022}. Most of the enterprises are implementing FIDO to reduce the attack surface and common human factors \cite{Alqubaisi_2020}. FIDO authentication has been very seamless \cite{Ghorbani_2020}. and user-friendly, including for differently abled people. It has been reshaping the user-authentication and digital identity scenario \cite{Campbell_Mark_2020} making it more reliable and safer from cyberattacks.

With device attestation, the question of migrating devices comes in, along with the question of authenticating from someone else’s devices. FIDO Passkeys \cite{XU_Sun_2021} might enable connecting one device to another over Bluetooth Low Energy (BLE) of device attestation \cite{Fernandez_2019}, pairing with QR Codes \cite{Lyastani_2022}. While this may claim to have proper device pairing and cryptographic exchanges, it is still possible for the attacker to socially engineer the target to scan the QR while being in the BLE range. To overcome all the above-mentioned problems, this paper proposes a novel approach for multi-device passwordless authentication.

In traditional cases, the user had to login to his account with other methods like legacy passwords and OTPs and enrol the new device. However, this was changed in September 2022, when Apple announced FIDO Passkeys. Passkeys allowed devices of a particular user to synchronize their identities so that the user can log in to apps and websites more securely across devices.

\subsection{TUSH-Key Advantages}

TUSH-Key (Transferable User Secrets on Hardware Key) is a seamless passwordless authentication enrolment system that enrols multiple devices of the same user for passwordless authentication seamlessly without synchronizing or cloning the private keys (secrets) among the devices. Passkeys, on the other hand, clone the secrets on multiple devices, in an end-to-end encrypted manner.

As per NIST SP 800-63B recommendation, “The key SHOULD be stored in suitably secure storage available to the authenticator application (e.g., keychain storage, TPM, TEE). The key SHALL be strongly protected against unauthorized disclosure using access controls that limit access to the key to only those software components on the device requiring access. Multi-factor cryptographic software authenticators SHOULD discourage and SHALL NOT facilitate the cloning of the secret key onto multiple devices” which passkeys often violate but TUSH-Key doesn’t.

TUSH-Key is aimed to provide all the features available to Passkeys but without violating standards like NIST, without relying on syncing parties like Google or Apple and making all transactions transparent.

\subsection{Primitives}

\begin{itemize}
    \item \textbf{User:} It is the entity that is trying to log in to an app or website from a device. The user is also known as the claimant.
    \item \textbf{Service Provider:} The Service Provider is an app or website where the user is attempting to authenticate to.
    \item \textbf{Relying Party (RP) Server:} This is a part of the service provider that FIDO authentication standard to verify the user. The RP Server is also known as the verifier.
    \item \textbf{Device:} This is the device that the user is trying to authenticate from. It is supposed to support FIDO2 specifications and have a secure storage media like a Trusted Platform Module (TPM) or a Trusted Execution Environment (TEE).
    \item \textbf{Browser:} This is a standard web browser that supports W3C WebAuthn and CTAP protocols. WebAuthn is the protocol used for communication between the RP Server and the browser, whereas CTAP is used for communication between the browser and the authenticator.
    \item \textbf{Authenticator:} This is the secure storage media like TPM or TEE with the device which performs the cryptographic operations. Authenticators are generally of two types: platform authenticator and cross-platform authenticator. Platform authenticators include the secure media which is inside the device (like TPM or TEE). Cross-platform authenticator includes external security keys or other devices that can work with the device over USB, NFC or BLE to store the secrets and perform cryptographic operations. It is to be noted that in certain cases a secondary device can also act as a cross-platform authenticator with the current device.
    \item \textbf{Challenge:} It is a random string of 16 bytes. It follows the challenge-response model to authenticate the claimant.
    \item \textbf{Keypair:} This is the RSA Keypair, which contains two parts, the private and the public keys. Usually, the private key is contained inside the authenticator. The public key, on the other hand, is sent to the RP Server.
    \item \textbf{Signing:} This involves applying a cryptographic operation with the private key on the challenge. This takes place inside the authenticator. A signed challenge can be verified by the RP server using the public key to verify the claimant.

\end{itemize}

\begin{figure}[ht!]
    \centering
    \includegraphics[width=0.8\textwidth]{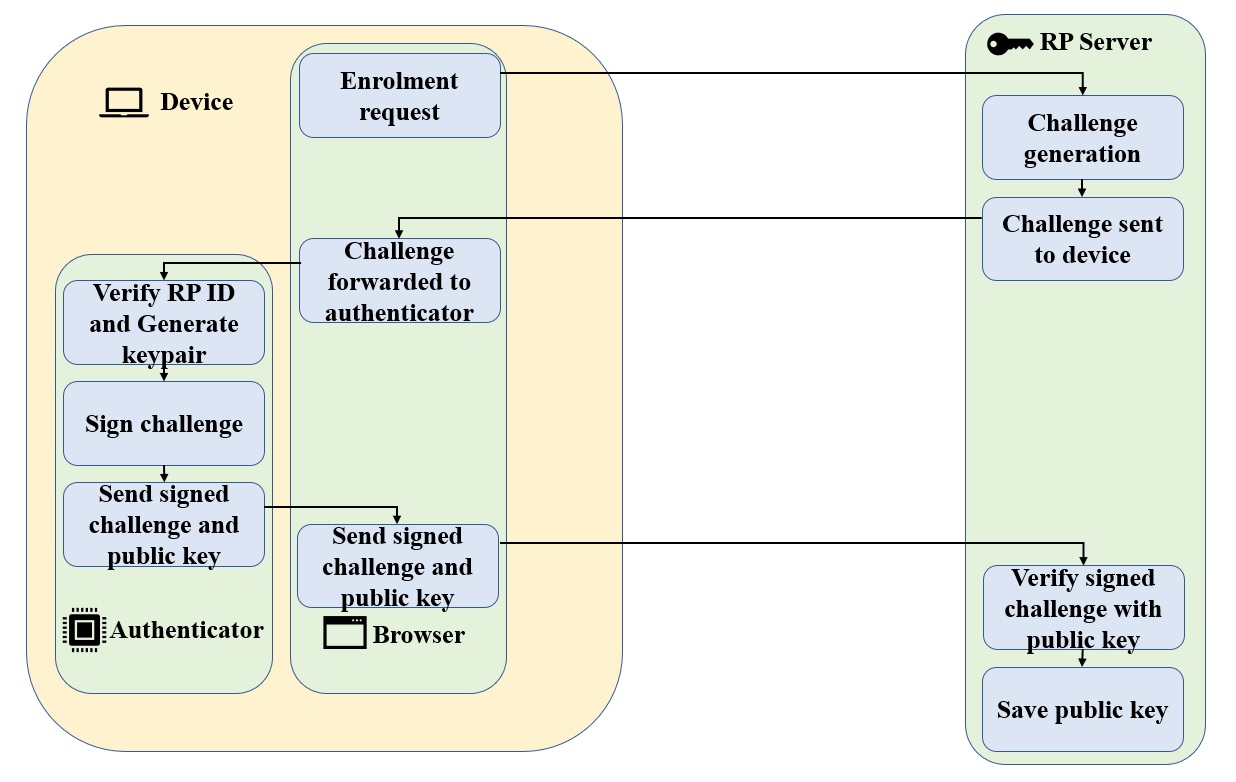}
    \caption{Device enrolment workflow.}
    \label{fig1}
\end{figure}

The architecture for traditional FIDO authentication includes the registration workflow and the authentication workflow. Figure \ref{fig1} shows the device enrolment or registration workflow as specified by FIDO Alliance.

\begin{figure}[ht!]
    \centering
    \includegraphics[width=0.8\textwidth]{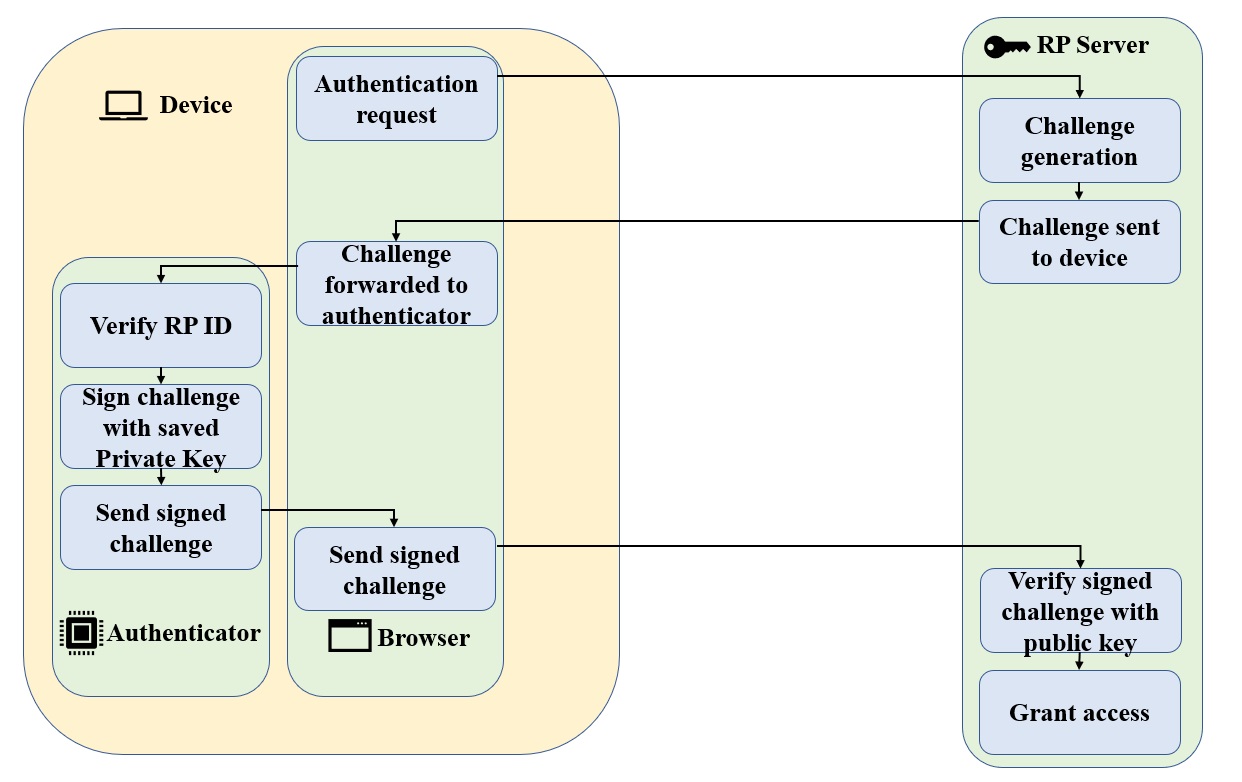}
    \caption{Authentication workflow.}
    \label{fig2}
\end{figure}

The registration workflow involves the browser to send an enrolment request to the RP Server. The RP Server would generate a 16-byte random challenge and return it to the browser. The browser then forwards the same to the Authenticator. The authenticator verifies the RP ID and then generates a new keypair and stores in the secure media. The challenge is signed with the private key. The signed challenge and the public key are sent to the RP Server. The RP Server verifies the signed challenge with the public key, and it saves the public key against the user’s account if the verification is successful. Further, it is to be noted that all communication between the device and the RP server is mandatorily secured with Transport Layer Security (TLS). This is to prevent any middleman attacks like Man in the Middle or Man in the Browser. Figure \ref{fig2} shows the authentication workflow.

The authentication workflow is like the registration workflow. The browser makes an authentication request to the RP Server. The RP Server generates a challenge and returns it to the browser, which then forwards it to the authenticator. The authenticator verifies the RP ID and then signs the challenge with the already saved private key. The RP Server verifies the same with the already saved public key and grants access to the user if the authentication process is successful.

It is to be noted that the private key never leaves the authenticator. But this limits the scope of device attestation since only one device is added to the user’s account. To further add a second device, the user must use legacy authentication methods on the second device and then enrol it. This makes it less secure as the legacy authentication methods are vulnerable.

To make this process easier, FIDO Alliance has come up with the idea of Passkeys. Passkeys are “a replacement for passwords that provide faster, easier, and more secure sign-ins to websites and apps across a user’s devices. Unlike passwords, passkeys are always strong and phishing resistant.” [FIDO Alliance website] It was first introduced by Apple in WWDC 2022 for MacOS and iOS devices. Later, it was also adopted by, Google for Android and ChromeOS.

\begin{figure}[ht!]
    \centering
    \includegraphics[width=0.8\textwidth]{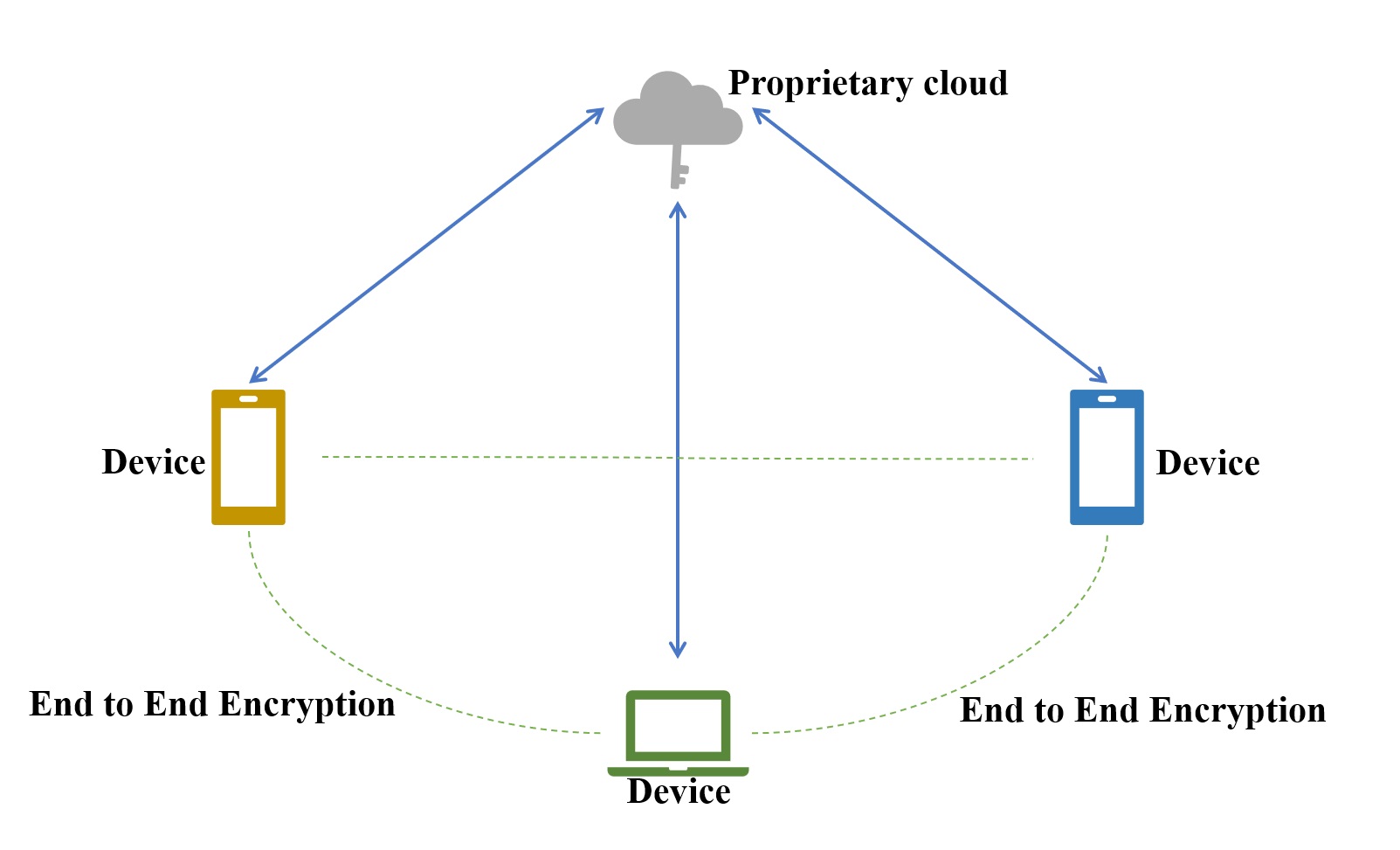}
    \caption{Architecture of Passkeys.}
    \label{fig3}
\end{figure}

Figure \ref{fig3} shows the architecture of passkeys. Passkeys synchronize the devices of a user by transferring the private keys from the Authenticator from one to the other. The communication between the devices is claimed to be end-to-end encrypted and they are synced over proprietary cloud services (iCloud Keychain for Apple devices and Google Password Manager for Android and ChromeOS). Further Apple allows sharing of passkeys over Airdrop so that you can share the private key to another device and use it on them without having to use the iCloud Keychain.

Though Passkeys make it very simple to share the private keys stored in authenticator, it has some serious implications. Passkeys mean a copy of the private keys are being stored in a proprietary cloud server (like iCloud Keychain). Further, the private keys are being cloned on other devices. 

This violates the NIST SP-800-63B \cite{Grassi_2017}. The clauses about Multi-Factor Cryptographic Software Authenticators include “The key SHOULD be stored in suitably secure storage available to the authenticator application (e.g., keychain storage, TPM, TEE). The key SHALL be strongly protected against unauthorized disclosure using access controls that limit access to the key to only those software components on the device requiring access. Multi-factor cryptographic software authenticators SHOULD discourage and SHALL NOT facilitate the cloning of the secret key onto multiple devices.” The cryptographic keys are stored in a proprietary cloud instead of or in addition to the secure storage available to the authenticator application. They main utility of passkeys is to clone the secret cryptographic key on multiple devices which is a direct violation of the above clause. This makes the use of passkeys vulnerable when used in critical systems.

\begin{figure}[ht!]
    \centering
    \includegraphics[width=0.8\textwidth]{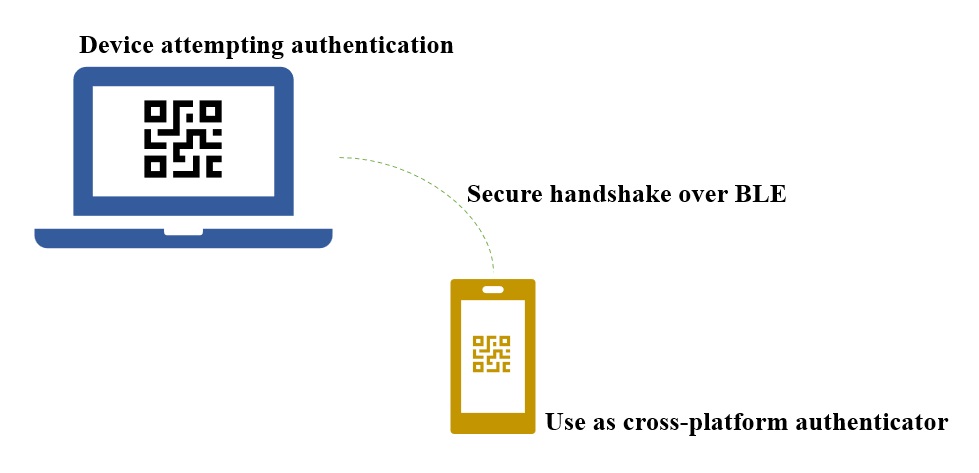}
    \caption{Use of QR codes.}
    \label{fig4}
\end{figure}

Further, when a passkey is to be used with a device which is not synchronized over the proprietary cloud, it involves using QR codes. The device trying to authenticate to the RP Server would display a QR Code. The device which is already enrolled with the RP Server would require to scan the QR Code and then both the devices would communicate securely over Bluetooth Low Energy (BLE) and the device already enrolled would act as cross-platform authenticator for the device trying to authenticate. Figure \ref{fig4} shows how QR codes are used for a secure handshake over BLE, followed by authentication.

The use of QR codes for a secure handshake among devices might be vulnerable to social engineering and insider attacks when the attacker is in the BLE range of the target. The attacker can socially engineer a legitimate user to scan a QR code and the user then grants access to the attacker.
To overcome the challenges faced above, the proposed solution aims at a more secure and seamless way of managing keys and secrets. The comparison of TUSH-Key and Passkeys are represented in Table \ref{tab2}.

\begin{table*}[htbp]
\caption{Comparison of TUSH-Key with Passkeys}
\begin{center}

\begin{tabular}{@{}lll@{}}
\hline
\textbf{Scope}                                                                                                                                                         & \textbf{Passkey}                                                                                                              & \textbf{TUSH-Key}                                                                                                                                 \\ \hline
\textbf{\begin{tabular}[c]{@{}l@{}}Cloning of secrets across a \\ user’s devices (violation of \\ NIST SP 800-63B)\end{tabular}}                                       & Yes                                                                                                                           & No                                                                                                                                                \\ \hline
\textbf{Storage of secrets}                                                                                                                                            & \begin{tabular}[c]{@{}l@{}}iCloud Keychain or Google 			\\ Password Manager (3rd party).\end{tabular}                         & On device TPM/TEE                                                                                                                                 \\ \hline
\textbf{\begin{tabular}[c]{@{}l@{}}Cross platform sync \\ (Windows to Android, \\ Android to iOS etc)\end{tabular}}                                                    & \begin{tabular}[c]{@{}l@{}}No. (Can use other devices 			\\ as external authenticator over \\ BLE with QR Codes)\end{tabular} & \begin{tabular}[c]{@{}l@{}}Automatic cross-platform \\ sync. No additional user \\ activity required.\end{tabular}                                \\ \hline
\textbf{\begin{tabular}[c]{@{}l@{}}Whether it possible to \\ register enrol a smartphone \\ for passwordless \\ authentication from a \\ Linux based PC.\end{tabular}} & No.                                                                                                                           & \begin{tabular}[c]{@{}l@{}}Yes, seamlessly with no \\ extra applications.\end{tabular}                                                            \\ \hline
\textbf{\begin{tabular}[c]{@{}l@{}}Whether enrolment of \\ other devices than the one \\ currently being used is \\ possible.\end{tabular}}                            & \begin{tabular}[c]{@{}l@{}}Yes, with QR Codes. But 			\\ phone needs to be in BLE \\ range.\end{tabular}                      & \begin{tabular}[c]{@{}l@{}}Yes, more seamlessly by just \\ logging into your TUSH-Key \\ account. Phone need not be in \\ BLE Range.\end{tabular} \\ \hline
\textbf{\begin{tabular}[c]{@{}l@{}}The keys being sharable \\ or exportable.\end{tabular}}                                                                             & \begin{tabular}[c]{@{}l@{}}Yes, private keys are 			\\ shared. Violation of NIST \\ SP 800-63B.\end{tabular}                  & \begin{tabular}[c]{@{}l@{}}Yes, but private keys not \\ shared.\end{tabular}                                                                      \\ \hline
\end{tabular}

\label{tab2}
\end{center}
\end{table*}

\section{TUSH-key: The Proposed Work}

TUSH-Key can be used for seamless and secure management of keys and secrets for passwordless authentication and device attestation. Once installed, TUSH-Key works with other devices of the user to enrol them against the user’s account with the RP Server. This enables the other devices of the user to seamlessly authenticate with the RP Server without the private keys being stored on proprietary clouds or the use of QR Codes and other authentication methods.
TUSH-Key works with devices on any platform and with compatible RP Servers.

\subsection{Primitives for TUSH-Key:}

\begin{itemize}
    \item \textbf{TUSH-Key Server:} It is the cloud server for communication with other devices of the User to facilitate performing cryptographic activities and key exchanges.
    \item \textbf{Symmetric key encryption:} This is a type of cryptographic algorithm where the same cryptographic key can be used to encrypt and decrypt a secret. When the encryption key is managed properly with key exchanges, it is can securely share a secret over an insecure channel. TUSH-Key uses AES-128 in Cipher Block Chaining (CBC) mode with the Fernet library.
    \item \textbf{Diffie-Hellman (DH) Key exchange:} It is the process of exchanging cryptographic keys over an insecure channel. It is a mathematical algorithm that generates a symmetric encryption key for two devices securely. It works by the two parties involved generating a pair of DH Keys and then they exchange the public keys. Applying a set of mathematical operations on the exchanged public keys and the retained private keys, it is possible to generate a secret which is common for both the parties performing the exchange. This shared secret can now be used as a key for symmetric encryption.
    \item \textbf{Daemon:} This is a background process running on the devices. This process is not under the control of any interactive user. It does not need any user interaction and is able to perform on its own. TUSH-Key Daemon manages the Diffie-Hellman key exchanges with the other devices of the user and enrol the device on with the RP Server against the user’s account.
    \item \textbf{Access token:} This is also referred to as ‘token’. This is a token that a RP Uses to identify which device belongs to which user.
    \item \textbf{Sender device:} The device that attempts to generate a TUSH-Key from the RP Server.
    \item \textbf{Receiver device:} Other devices of the user except the Sender device. Both the sender and the receiver devices are assumed to be running the Daemon.
    
\end{itemize}

When the user requests the RP Servers to generate a TUSH-Key, the RP Server generates an Access Token for the user. This access token is then sent to the current device the user is on. The device forwards the same to the daemon. The daemon performs Diffie-Hellman key exchange with the other devices using the TUSH-Key Server.

\textbf{TUSH-Key has two flows:}

\begin{itemize}
    \item \textbf{Device Registration flow:} 
    
    Here the user installs the TUSH-Key Daemon on the system. Device Registration flow when the TUSH-Key daemon is run for the first time on a system. TUSH-Key relies on the Microsoft SSO for verification of the user identity. It takes the user to the Microsoft SSO Page where the user signs in. TUSH-Key then uses the Microsoft OAuth2 workflow to retrieve the email id of the user, which becomes the user id for the user. Figure \ref{fig5} shows the SSO permission page.

    \begin{figure}[htbp]
        \centering
        \includegraphics[width=0.5\textwidth]{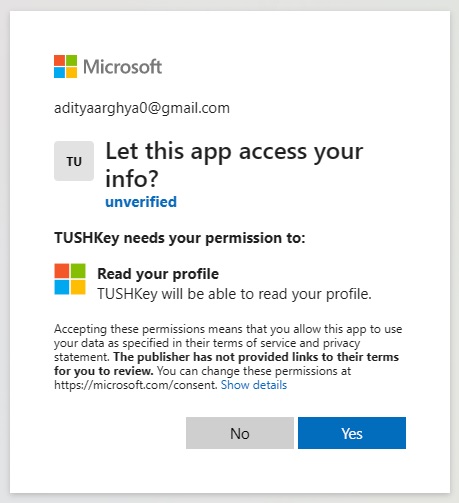}
        \caption{SSO Permission page.}
        \label{fig5}
    \end{figure}

    The Daemon then generates a unique device id which is of the RFC 4122 UUID version 4 type. Further, it generates a Diffie-Hellman keypair. The Diffie-Hellman keypair and the device ID is stored on the disk of the system.
    
    The Diffie-Hellman public key, the device ID and the user ID is then sent to the TUSH-Key Server by an API Call. The TUSH-Key server stores them for further operations. Figure \ref{fig6} shows the TUSH-Key device registration flow.

    \begin{figure}[htbp]
        \centering
        \includegraphics[width=0.8\textwidth]{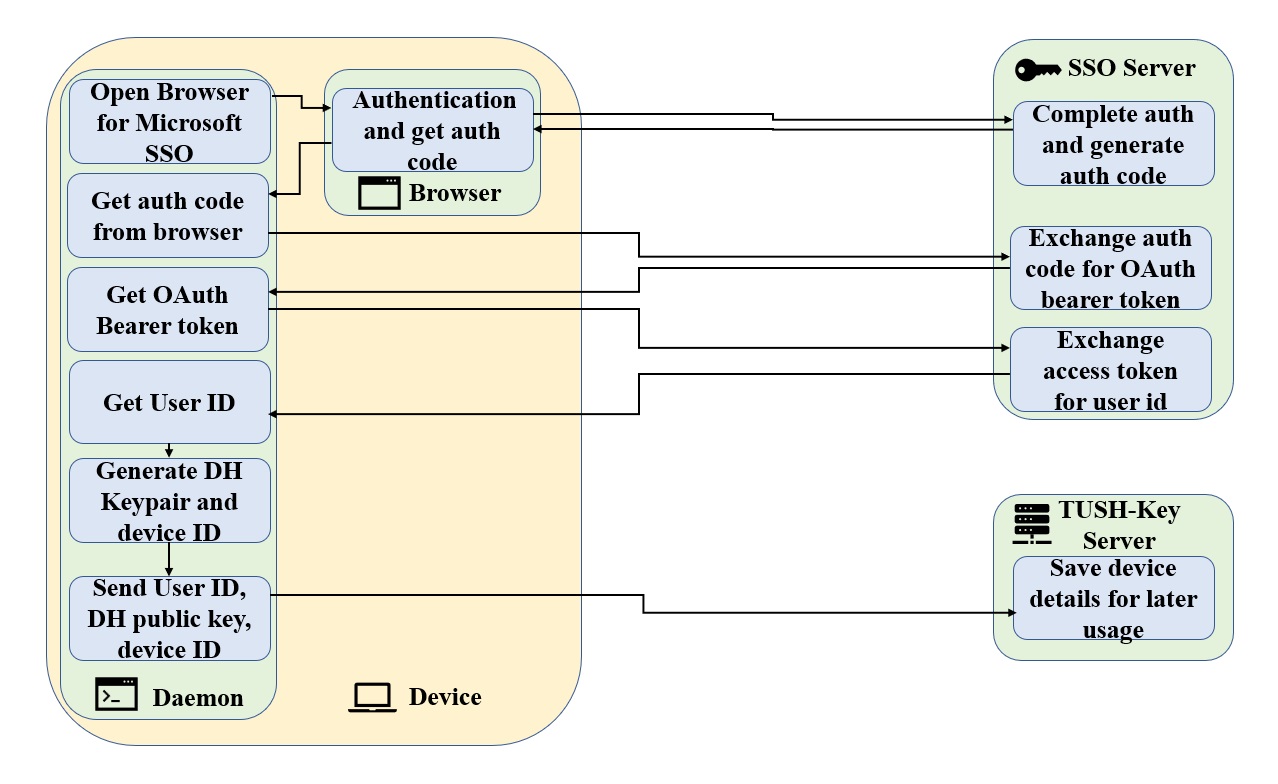}
        \caption{TUSH-Key Device Registration flow.}
        \label{fig6}
    \end{figure}

    \item \textbf{TUSH-Key Sync flow:} 

    This flow takes care of enrolling the other devices of the user with the RP Server when one device is enrolled. This works by the sender device and the receiver device(s) working together to enrol the receiver devices. 

    When the sender device wants to generate TUSH-Key for the other devices of the user, the sender device requests the RP to generate an access token. This access token can be used to login to the user’s account without the need of explicit authentication and this automatically registers the device logging in with this access token for passwordless authentication for the user. The sender device needs to send this token to the receiver device(s) in a seamless manner securely over an insecure channel. 

    Now, the receiver device is assumed to already be registered with the TUSH-Key server. This implies that the device id, and the Diffie-Hellman public key of the receiver device is already with TUSH-Key server. The sender device requests the TUSH-Key server with its own device id. TUSH-Key server retrieves the user id using the sender device id. It then retrieves the other devices of the same user and their Diffie Hellman Public Keys. The same is returned to the sender device.

    The sender device now performs generates a shared secret using its own Diffie-Hellman key and the Diffie-Hellman public key of the other device. Once the shared secret is generated, the access token is encrypted by symmetric encryption (AES-128) with the shared secret as the encryption key. The encrypted access token and the respective sender and receiver device ids are sent to the TUSH-Key server. Figure \ref{fig7} shows the TUSH-Key sync flow on the sender side.

    \begin{figure}[htbp]
        \centering
        \includegraphics[width=0.8\textwidth]{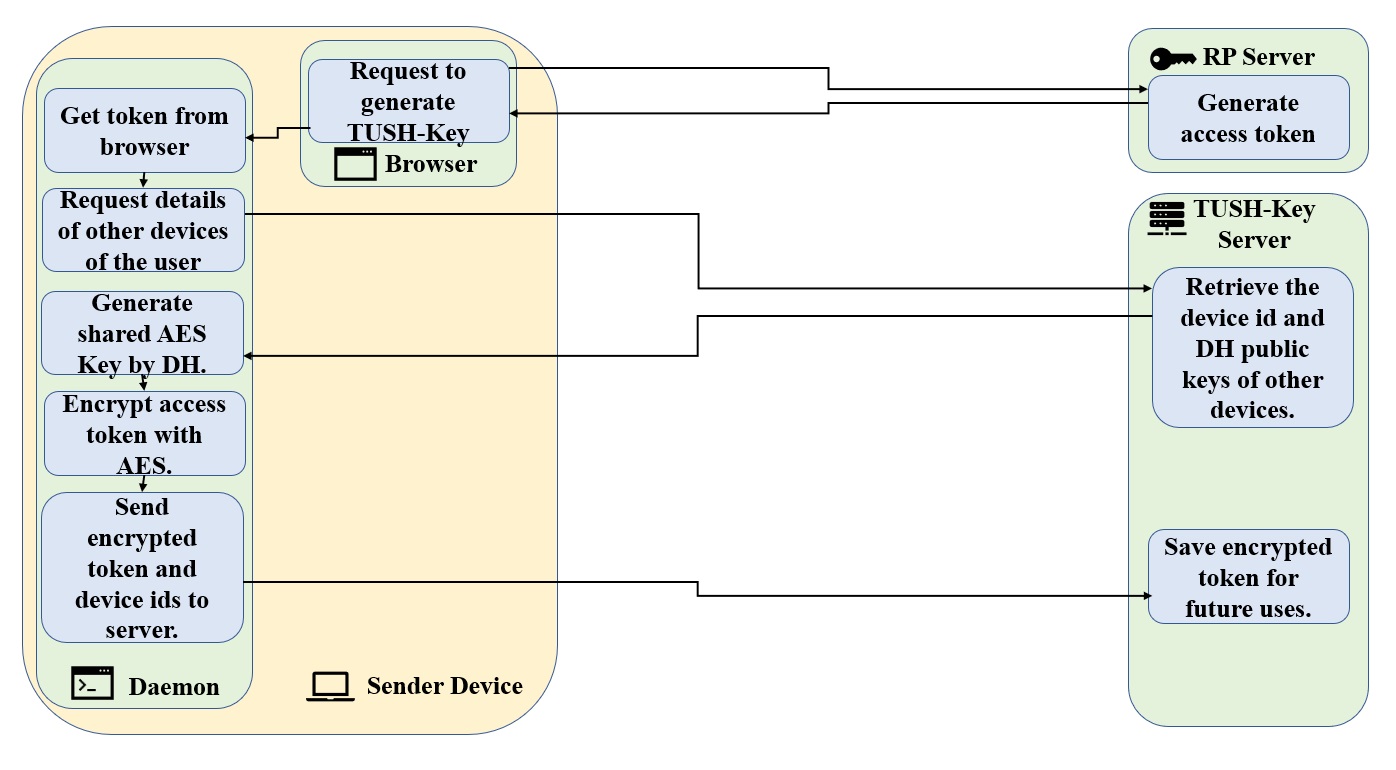}
        \caption{TUSH-Key sync flow - Sender side.}
        \label{fig7}
    \end{figure}

    The receiver device daemon requests the TUSH-Key server for pending encrypted tokens with the receiver device id continuously at regular intervals. The TUSH-Key, after receiving the encrypted token from the sender device responds to the receiver device. The receiver device requests for the encrypted token and the sender device id. It then requests the TUSH-Key server for the DH Public key of the sender device. On getting the DH Public key, it computes the shared secret key using its own DH Keys. Once the shared secret is obtained, the encrypted access token is decrypted with AES. The receiver then uses the access token and requests the RP Server to enrol the device for passwordless authentication against the user’s account. Figure \ref{fig8} shows the TUSH-Key Sync flow on receiver side.

    \begin{figure}[htbp]
        \centering
        \includegraphics[width=0.8\textwidth]{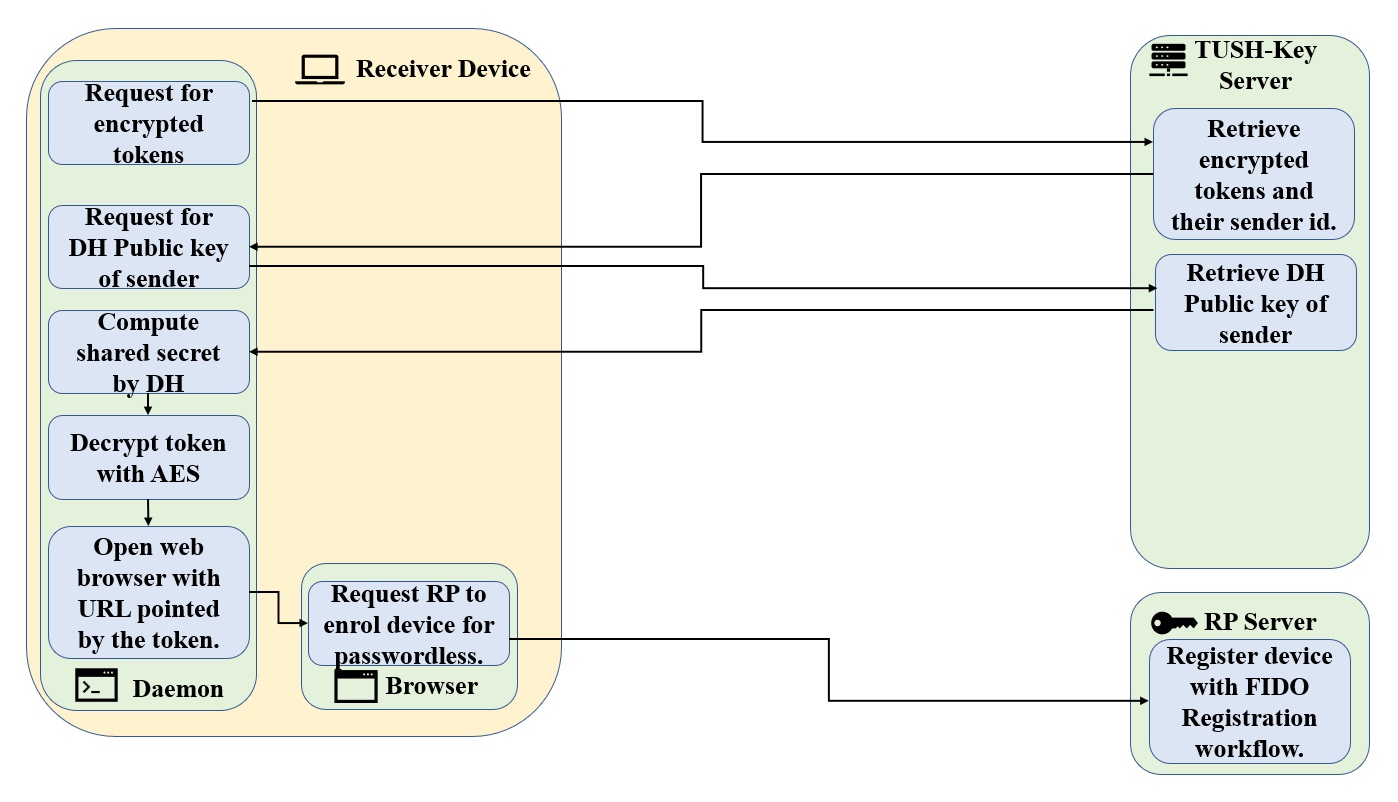}
        \caption{TUSH-Key sync flow - Receiver side.}
        \label{fig8}
    \end{figure} 
    
\end{itemize}

Once the receiver device is enrolled for passwordless authentication with the RP Server, the user can use it to seamlessly authenticate to the service provider. It is to be noted that TUSH-Key does not clone the authenticator secret of one device to another. Instead, the other devices generate their new authenticator secrets, and all the public keys are stored with the RP. Hence, the devices do not facilitate cloning of the cryptographic secrets. The secrets are stored in the secure media available to the authenticator itself and not in any proprietary cloud server. This complies with the NIST SP 800-63B. These comply with the NIST SP 800-63B.

\subsection{Experimental Validation of TUSH-Key}

\begin{itemize}
    \item \textbf{Experimental setup:} The test RP Server and the TUSH-Key server have been implemented using Azure B1s Virtual Machines running Ubuntu 18.04 LTS. They both have a SQL Server each connected to them. The cryptographic public keys are stored on BLOB storages connected to each of the servers. The devices that have tested as sender and receiver devices are two Windows PCs and two Android devices in all combinations. The firewalls for the cloud servers have been configured to protect against malicious users.
    \item \textbf{Performance analysis:} It has been observed that TUSH-Key Sync flow takes very less time to enrol the devices for passwordless authentication, plus user input times. For the tests, all devices had internet connectivity of about 150 Mbps speed. Hence, it can be claimed that TUSH-Key is quite efficient in the experimental setup and can be more efficient with better servers and connectivity.

    The average time taken for the TUSH-Key Sync flow is 447 ms. Further, the average time taken for enrolling a device with FIDO2 specifications under the experimental setup is about 580 ms. Hence, the total time needed for automatically enrolling the other devices of the user against the user’s account in the RP Server is roughly 1 sec. Table \ref{tab3} shows the time taken for TUSH-Key sync on different platforms.

    \begin{table*}[htbp]
    \caption{TUSH-Key sync time}
    \begin{center}
    \begin{tabular}{@{}llll@{}}
    \hline
    Sl. No. & Sender Device                                                              & Receiver Device                                                          & \begin{tabular}[c]{@{}l@{}}Time for TUSH-Key \\ Sync flow\end{tabular} \\ \hline
    1       & \begin{tabular}[c]{@{}l@{}}Windows PC \\ (Dell Inspiron 3593)\end{tabular} & \begin{tabular}[c]{@{}l@{}}Windows PC\\ (Galaxy Book 3 Pro)\end{tabular} & 436 ms                                                                 \\
    2       & \begin{tabular}[c]{@{}l@{}}Windows PC\\ (Galaxy Book 3 Pro)\end{tabular}   & \begin{tabular}[c]{@{}l@{}}Android\\ (Galaxy Z Flip 4)\end{tabular}      & 462 ms                                                                 \\
    3       & \begin{tabular}[c]{@{}l@{}}Android\\ (Galaxy M51)\end{tabular}             & \begin{tabular}[c]{@{}l@{}}Android\\ (Galaxy Z Flip 4)\end{tabular}      & 443 ms                                                                 \\ \hline
    \multicolumn{3}{c}{Average}                                                                                                                                     & 447 ms      \\                         \hline                                 
    \end{tabular}

    \label{tab3}
    \end{center}
    \end{table*}

    Table \ref{tab4} shows the time taken for enrolling the receiver device with RP Server following the FIDO Specifications.

    \begin{table*}[htbp]
    \caption{Device enrolment time}
    \begin{center}
    \begin{tabular}{@{}llll@{}}
    \hline 
    Sl. No. & \begin{tabular}[c]{@{}l@{}}Time taken to create \\ challenge by RP\end{tabular} & \begin{tabular}[c]{@{}l@{}}Time taken to generate \\ keypair, sign challenge \\ and verify\end{tabular} & \begin{tabular}[c]{@{}l@{}}Total time taken to \\ enrol device with \\ RP Server\end{tabular} \\ \hline 
    1       & 302 ms                                                                          & 288 ms                                                                                                  & 590 ms                                                                                        \\
    2       & 311 ms                                                                          & 277 ms                                                                                                  & 588 ms                                                                                        \\
    3       & 311 ms                                                                          & 64 ms                                                                                                   & 375 ms                                                                                        \\
    4       & 412 ms                                                                          & 312 ms                                                                                                  & 724 ms                                                                                        \\
    5       & 346 ms                                                                          & 274 ms                                                                                                  & 620 ms                                                                                        \\ \hline
    \multicolumn{3}{c}{Average}                                                                                                                                                                         & 580 ms    \\                         \hline                                                                                     
    \end{tabular}

    \label{tab4}
    \end{center}
    \end{table*}
    
\end{itemize}

The TUSH-Key Server virtual machine uses elastic scalability, which allows it to scale up or down as necessary. To evenly distribute the load among all the VMs in the VM set, it is further processed by an elastic load balancer. The backend is robust to most disruptions thanks to availability sets and availability zones. The monthly uptime should exceed 99.99\% per the Service Level Agreements (SLA), and employing Availability Sets, Availability Zones can be helpful in maintaining the service even in the event of a disaster or catastrophic failure in a data centre.

\section{Security Analysis using AVISPA}

Automated Validation of Internet Security Protocols and Applications (AVISPA) is a suite widely used for examining and validating security protocols \cite{DAS_Sibi_2022}. AVISPA offers a set of frameworks enabling the formal evaluation of security attributes in a variety of protocols, including network protocols, online security protocols, and cryptographic protocols. To make sure that these protocols are secure and proper, AVISPA integrates a number of automated analysis approaches. For modelling and analysing security protocols, AVISPA makes use of the language High-Level Protocol Specification Language (HLPSL). The protocol's messages, responsibilities, and security attributes are captured in a clear, formal representation provided by HLPSL. Using a variety of analysis approaches, the primary analysis engine Automatic Protocol Analyzer (APPA) instantly analyses the protocol specification specified in HLPSL to look for security flaws and property violations. HLPSL advantages from declarative semantics based on Temporal Logic of Actions in addition to functional semantics depending on Intermediate Format (IF). The HLPSL2IF converter converts HLPSL specifications into matching IF specifications. The AVISPA Tool's back-ends take IF specifications and execute them using a number of analytical techniques. At present, AVISPA supports the following four back-ends: On-the-fly Model-Checker (OFMC), Constraint-Logic-based Attack Searcher (CL-AtSe), SAT-based Model-Checker (SATMC), and TA4SP (Tree Automata based on Automatic Approximations for the Analysis of Security Protocols).

The proposed TUSHKEY is successfully validated with OFMC and ATSE back-ends. Firstly, the Symbolic model checker OFMC examines the TUSHKEY protocol implementation and confirms particular security attributes to ensure that the security protocols are appropriate. By using a demand-driven strategy within the transition system with the assistance of IF specifications, OFMC is able to carry out protocol deception and constrained verification \cite{DAS_Sibi_2022}. Secondly, an automated theorem proving methods such as CL-ATSE theorem prover is employed to demonstrate security features. Utilizing strong simplification and redundancy-removal techniques, CL-ATSE solves problems under constraints. CL-ATSE helps in detecting potential security weaknesses, vulnerabilities, or design flaws in protocols and applications, thus ensuring their safety and preventing potential attacks.

\begin{figure*}[htbp]
    \centering
    \includegraphics[width=0.8\textwidth]{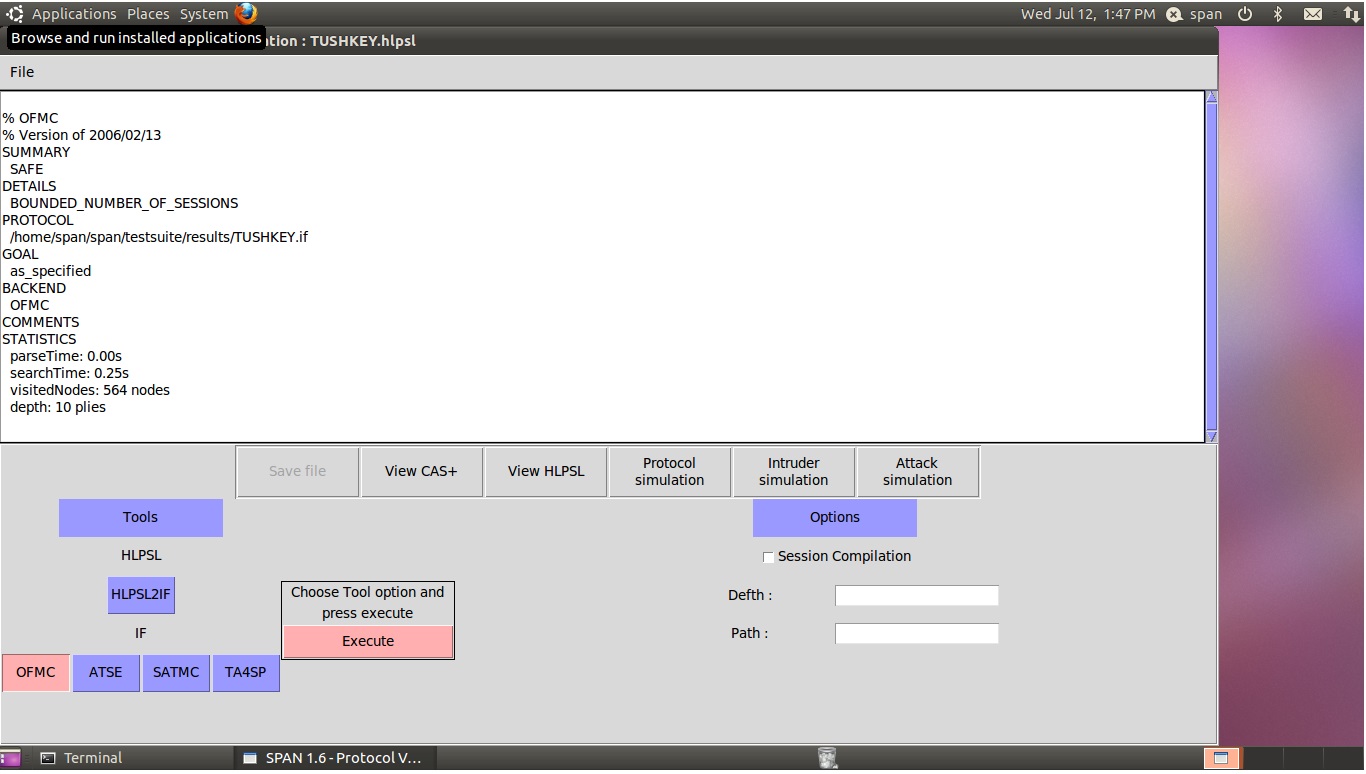}
    \caption{Security analysis results of TUSHKEY using AVISPA (OFMC backend).}
    \label{fig9}
\end{figure*}

\begin{figure*}[htbp]
    \centering
    \includegraphics[width=0.8\textwidth]{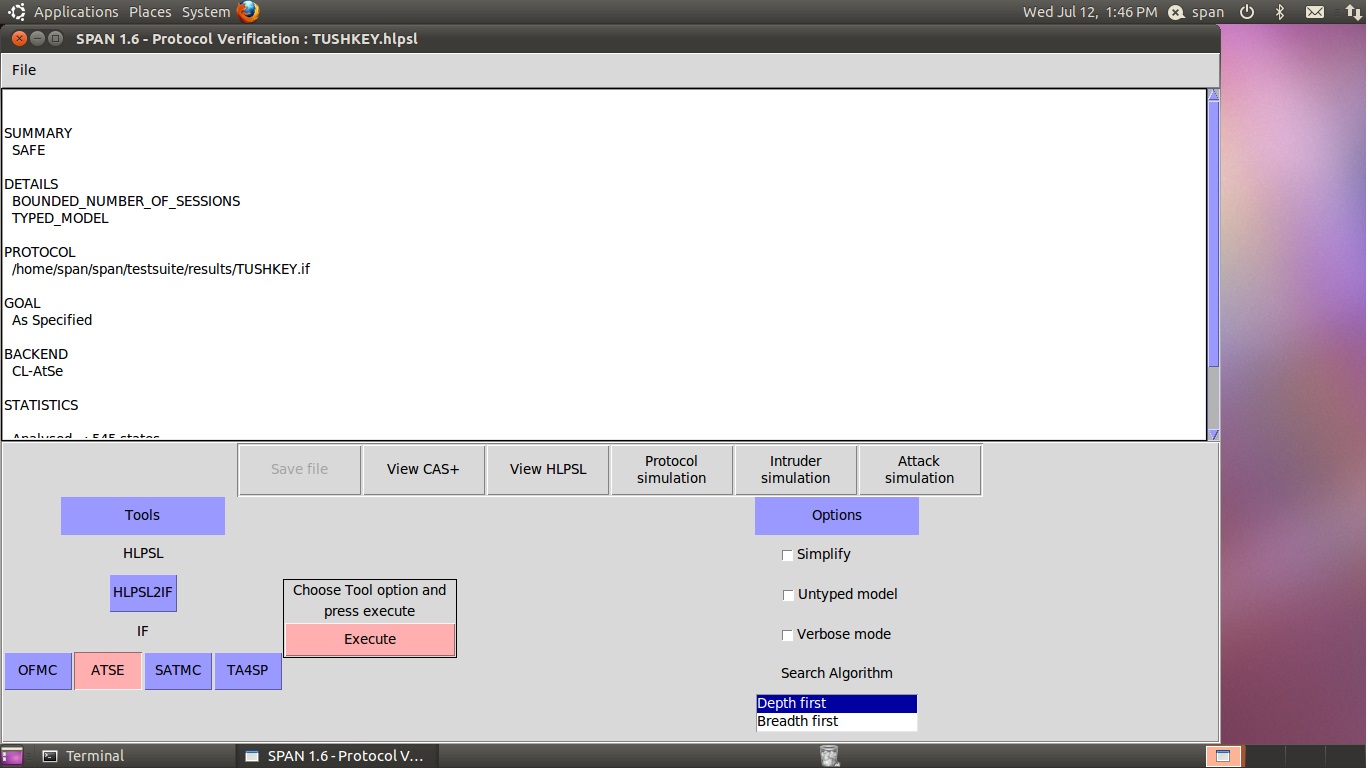}
    \caption{Security analysis results of TUSHKEY using AVISPA (ATSE backend).}
    \label{fig10}
\end{figure*}

Finally, the analysis of the proposed TUSHKEY using AVISPA reveals that both the OFMC (Figure \ref{fig9}) and ATSE (Figure \ref{fig10}) back ends have produced results indicating that the protocol is deemed safe. The safety of the protocol has been independently verified through the utilization of exhaustive model checking techniques with OFMC and formal methods for validation with ATSE. These results provide a strong level of assurance, indicating that TUSHKEY has undergone a thorough examination for potential vulnerabilities, meeting the necessary safety standards.

\section{Conclusion and Future Scope}

This paper presents a standard that allows enrolling all devices of a particular user to service providers for passwordless authentication. This makes the use of multiple devices more seamless without any lapse in security. There is proper cryptographic key management that ensures the private keys are secure and stays on secure storage media on the device itself and not cloned onto other devices via in-secure channels or proprietary clouds. This complies with standards like the NIST SP 800-63B and can be implemented in sensitive scenarios as well. A reference implementation has been shown with test devices and have reached desired outputs. This will be further extended for iOS, iPad OS to make it truly platform independent and without relying on proprietary services.

\bibliographystyle{IEEEtran} 
\bibliography{arXiv_2023-XXX-R0_TUSHKEY-1.bib}

\begin{thebibliography}{10}
\providecommand{\url}[1]{#1}
\csname url@samestyle\endcsname
\providecommand{\newblock}{\relax}
\providecommand{\bibinfo}[2]{#2}
\providecommand{\BIBentrySTDinterwordspacing}{\spaceskip=0pt\relax}
\providecommand{\BIBentryALTinterwordstretchfactor}{4}
\providecommand{\BIBentryALTinterwordspacing}{\spaceskip=\fontdimen2\font plus
\BIBentryALTinterwordstretchfactor\fontdimen3\font minus
  \fontdimen4\font\relax}
\providecommand{\BIBforeignlanguage}[2]{{%
\expandafter\ifx\csname l@#1\endcsname\relax
\typeout{** WARNING: IEEEtran.bst: No hyphenation pattern has been}%
\typeout{** loaded for the language `#1'. Using the pattern for}%
\typeout{** the default language instead.}%
\else
\language=\csname l@#1\endcsname
\fi
#2}}
\providecommand{\BIBdecl}{\relax}
\BIBdecl

\bibitem{Fano_1966}
\BIBentryALTinterwordspacing
F.~R. M. and C.~F. J., ``Time sharing on computers,'' p. 128–143, 1966, last
  accessed on 12.07.2023. [Online]. Available:
  \url{http://simson.net/ref/lcs_35/1966_ScientificAmerican_MAC.pdf}
\BIBentrySTDinterwordspacing

\bibitem{Al-Kabir_2022}
M.~A. Al~Kabir and W.~Elmedany, ``An overview of the present and future of user
  authentication,'' in \emph{Proc. Intl. Conf. Mid. East. Nor. Afr. Comm.},
  2022, pp. 10--17.

\bibitem{Huseynov_2022}
E.~Huseynov, ``Passwordless vpn using fido2 security keys: Modern
  authentication security for legacy vpn systems,'' in \emph{Proc. Intl. Conf.
  Data. Intel. Sec.}, 2022, pp. 01--03.

\bibitem{KINGO_2023}
T.~Kingo and D.~F. Aranha, ``User-centric security analysis of mitid: The
  danish passwordless digital identity solution,'' \emph{Computers \&
  Security}, vol. 132, p. 103376, 2023.

\bibitem{Bicakci_2022}
K.~Bicakci and Y.~Uzunay, ``Is fido2 passwordless authentication a hype or for
  real?: A position paper,'' in \emph{Proc. Intl. Conf. Info. Sec. Crypt.},
  2022, pp. 68--73.

\bibitem{Laing_2022}
T.~Laing, E.~Marin, M.~D. Ryan, J.~Schiffman, and G.~Wattiau, ``Symbolon:
  Enabling flexible multi-device-based user authentication,'' in \emph{Proc.
  Intl. Conf. Dep. Sec. Comp.}, 2022, pp. 1--12.

\bibitem{Liou_2021}
W.-C. Liou and T.~Lin, ``T-auth: A novel authentication mechanism for the iot
  based on smart contracts and pufs,'' in \emph{Proc. Intl. Conf. Comm. Work.},
  2021, pp. 1--6.

\bibitem{Kaczmarek_Ozturk_Tsudik_2019}
T.~Kaczmarek, E.~Ozturk, and G.~Tsudik, ``Thermanator,'' 2019, p. 586–593.

\bibitem{Parmar_2022}
V.~Parmar, H.~A. Sanghvi, R.~H. Patel, and A.~S. Pandya, ``A comprehensive
  study on passwordless authentication,'' in \emph{Proc. Intl. Conf. Sus. Comp.
  Data Comm. Sys.}, 2022, pp. 1266--1275.

\bibitem{Arias_2018}
O.~Arias, F.~Rahman, M.~Tehranipoor, and Y.~Jin, ``Device attestation: Past,
  present, and future,'' in \emph{Proc. Intl. Conf. Des. Auto. Test. Exh.},
  2018, pp. 473--478.

\bibitem{Morii_2017}
M.~Morii, H.~Tanioka, K.~Ohira, M.~Sano, Y.~Seki, K.~Matsuura, and T.~Ueta,
  ``Research on integrated authentication using passwordless authentication
  method,'' in \emph{Proc. Intl. Conf. Comp. Soft. App.}, vol.~1, 2017, pp.
  682--685.

\bibitem{Hodges_2021}
\BIBentryALTinterwordspacing
J.~Hodges, J.~Jones, M.~B. Jones, A.~Kumar, and E.~Lundberg, ``Web
  authentication: An api for accessing public key credentials level 2,'' 2021,
  last accessed on 12.07.2023. [Online]. Available:
  \url{https://www.w3.org/TR/webauthn-2/}
\BIBentrySTDinterwordspacing

\bibitem{Campbell_2023}
M.~Campbell, ``The road to decentralized identity: The techniques, promises,
  and challenges of tomorrow’s digital identity,'' \emph{Computer}, vol.~56,
  no.~6, pp. 96--100, 2023.

\bibitem{Sibi_2022}
S.~C. Sethuraman, A.~Mitra, G.~Galada, A.~Ghosh, and S.~Anitha, ``Metakey: A
  novel and seamless passwordless multifactor authentication for metaverse,''
  in \emph{Proc. Intl. Symp. Sma. Elec. Sys.}, 2022, pp. 662--664.

\bibitem{Sibi_Loki_2022}
S.~Chakkaravarthy~Sethuraman, A.~Mitra, K.-C. Li, A.~Ghosh, M.~Gopinath, and
  N.~Sukhija, ``Loki: A physical security key compatible iot based lock for
  protecting physical assets,'' \emph{IEEE Access}, vol.~10, pp.
  112\,721--112\,730, 2022.

\bibitem{Alqubaisi_2020}
F.~Alqubaisi, A.~S. Wazan, L.~Ahmad, and D.~W. Chadwick, ``Should we rush to
  implement password-less single factor fido2 based authentication?'' in
  \emph{Intl. Conf. Und. Res. Appl. Comp.}, 2020, pp. 1--6.

\bibitem{Ghorbani_2020}
S.~Ghorbani~Lyastani, M.~Schilling, M.~Neumayr, M.~Backes, and S.~Bugiel, ``Is
  fido2 the kingslayer of user authentication? a comparative usability study of
  fido2 passwordless authentication,'' in \emph{Intl. Symp. Sec. Priv.}, 2020,
  pp. 268--285.

\bibitem{Campbell_Mark_2020}
M.~Campbell, ``Putting the passe into passwords: How passwordless technologies
  are reshaping digital identity,'' \emph{Computer}, vol.~53, no.~8, pp.
  89--93, 2020.

\bibitem{XU_Sun_2021}
P.~Xu, R.~Sun, W.~Wang, T.~Chen, Y.~Zheng, and H.~Jin, ``Sdd: A trusted display
  of fido2 transaction confirmation without trusted execution environment,''
  \emph{Future Generation Computer Systems}, vol. 125, pp. 32--40, 2021.

\bibitem{Fernandez_2019}
F.~Bascunana and Gema, ``Validation of authentication measures implementation
  in iot mobile applications,'' \emph{Smart Cities}, vol.~2, no.~2, pp.
  163--178, 2019.

\bibitem{Lyastani_2022}
S.~G. Lyastani, M.~Backes, and S.~Bugiel, ``A systematic study of the
  consistency of two-factor authentication user journeys on top-ranked websites
  (extended version),'' 2022.

\bibitem{Grassi_2017}
\BIBentryALTinterwordspacing
P.~A. Grassi, M.~E. Garcia, and J.~L. Fenton, ``Digital identity guidelines:
  authentication and lifecycle management,'' pp. 3--63, 2017, last accessed on
  12.07.2023. [Online]. Available:
  \url{https://oag.ca.gov/sites/all/files/agweb/pdfs/bciis/03-nist-sp-800-63-3.pdf}
\BIBentrySTDinterwordspacing

\bibitem{DAS_Sibi_2022}
D.~Das, S.~C. Sethuraman, and S.~C. Satapathy, ``A decentralized open web
  cryptographic standard,'' \emph{Computers and Electrical Engineering},
  vol.~99, p. 107751, 2022.

\end{thebibliography}

\end{document}